\documentclass[11pt,a4paper]{article}
\pdfoutput=1

\usepackage{jheppub}
\usepackage{dcolumn}
\usepackage{bm}
\usepackage{mathtools}
\usepackage{slashed}
\usepackage{xfrac}
\usepackage{booktabs}
\usepackage{tabularx}
\usepackage{array}
\usepackage{comment}
\usepackage{physics}
\usepackage{enumitem}
\usepackage{placeins}
\usepackage{capt-of}
\usepackage{float}
\usepackage{subcaption}
\usepackage{adjustbox}
\usepackage{xcolor}

\usepackage{tikz}
\usetikzlibrary{arrows.meta,decorations.pathmorphing}

\renewcommand{\arraystretch}{1.25}

\newcommand{\CF}{\mathcal{F}_{\rm C}}

\newcolumntype{Y}{>{\centering\arraybackslash}X}

\newenvironment{revision}{\begingroup\color{black}}{\endgroup}

\preprint{CHIBA-EP-282}

\title{Genuine Tripartite Entanglement Selects Gauge-Invariant Theories}
\author{Junya Yamagishi}
\emailAdd{junya@chiba-u.jp}
\affiliation{Department of Physics, Graduate School of Science, Chiba University, Chiba 263-8522, Japan}

\abstract{
Recent studies have explored whether physical constants and symmetries can be selected by extremizing bipartite entanglement entropy. We extend this approach to a three-particle setting. Specifically, we consider a $2+1$-particle system in which particles $B$ and $C$ are initially entangled, and then particles $A$ and $B$ scatter without the direct participation of particle $C$ for both gluon-gluon and graviton-graviton scatterings.
Using angular expansions around the forward and backward scattering limits, we analyze bipartite entanglement in the corresponding two-particle system of $A$ and $B$ and two measures of genuine tripartite entanglement (GTE), the concurrence fill and the generalized geometric measure (GGM), in the three-particle system. As is already known, the bipartite entanglement provides no universal criterion for selecting the symmetry-preserving theory: depending on the initial helicities, it may either maximize or minimize the entanglement entropy.
By contrast, we find that the GTE-based analysis uniquely identifies the
gauge-invariant and diffeomorphism-invariant theories as those that suppress
the GTE in gluon-gluon and graviton-graviton scatterings, respectively. This result suggests a nontrivial connection between tripartite entanglement and the gauge and
diffeomorphism symmetries of fundamental interactions.
}

\begin{document}

\makeatletter\renewcommand{\@fpheader}{\ }\makeatother
\maketitle
\renewcommand{\thefootnote}{\#\arabic{footnote}}
\setcounter{footnote}{0}
\flushbottom

\section{Introduction}
\label{sec:introduction}
Recent work has investigated whether entanglement entropy generated in quantum-field-theoretic scattering can reveal the coupling structure and symmetries of interactions.
Because scattering amplitudes can generate quantum correlations among the internal degrees of freedom of final-state particles, final-state entanglement can serve as an information-theoretic probe of interactions.
Requiring the final-state entanglement to be maximal has been used to account for physical constants and symmetry structures of the Standard Model and its extensions~\cite{CerveraLierta2017,CarenaMirror2025,Liu2026Higgs,Li2026Composite}.
Conversely, the suppression of the final-state entanglement has also been related to symmetry structures and flavor patterns~\cite{Beane2018,Low2021,Liu2021,Carena2024,LiuLow2024,Hu2024,Chang2024,Kowalska2024,Thaler2025,Hu2025}.
Nevertheless, no unified interpretation of these results has emerged, and the appropriate information-theoretic selection principle remains unclear.
Alongside these entanglement-based analyses, increasing attention has recently been paid to the magic, or nonstabilizerness, of quantum states in nuclear and particle physics, as quantified by the stabilizer R\'enyi entropy (SRE)~\cite{Leone2022,Robin2025,Chernyshev2025,White2024,LiuMaximal2026,Liu2025,Aoude2025,Busoni2025,Gargalionis2026,Chu2026}.

Throughout this work, we use ``gauge invariance'' as a generic term covering both the gauge theory and the gravitational theory considered here; for gravity, it should be understood specifically as diffeomorphism invariance.
Following Refs.~\cite{Cao2026,Nunez2026Gauge}, we consider a family of theories in which a real parameter $k$ deforms the relative coefficient of the four-point vertex, as specified in Sec.~\ref{sec:amplitudes}; at $k=1$ the undeformed, gauge-invariant theory is recovered.
Reference~\cite{Nunez2026Gauge} found that entanglement extrema alone do not single out this theory.
However, when the entanglement-extremization criterion is supplemented by the minimization of magic, as quantified by the SRE, it is uniquely selected in gluon-gluon scattering and graviton-graviton scattering.
Minimizing the SRE also yields a value close to the weak mixing angle~\cite{Liu2025-2}.
These findings motivate the study of additional information-theoretic criteria.

Tripartite quantum entanglement has also recently attracted increasing attention in collider physics, with studies of three-body final states~\cite{Sakurai2024,Horodecki2025,Goncalves2026,Banacki2026}.
In particular, a $2+1$-particle system provides a framework for probing the redistribution of scattering-induced correlations in greater detail.
In this setup, particles $B$ and $C$ are initially entangled, and then particles $A$ and $B$ scatter without the direct participation of particle $C$.
In this paper, we assume that particle $C$ remains spatially separated from $A$ and $B$ throughout the process.
The setup was first proposed in Ref.~\cite{Araujo2019} and has since been used to study information encoded in the state of particle $C$, the transfer of bipartite entanglement, and the generation of genuine tripartite entanglement (GTE) in QED scattering~\cite{Fonseca2022,Blasone2024,Cao2026}.

However, tripartite correlations of this kind have not previously been used to select the gauge-invariant theory among a family of deformed ones.
We therefore investigate whether tripartite correlations involving a spectator particle can select this theory, which extrema of bipartite entanglement alone fail to select uniquely.
More specifically, we study a $2+1$-particle extension of the deformed theory introduced above.
We then quantify how scattering redistributes the initial $BC$ entanglement into the GTE.
We focus on the forward and backward scattering limits, where the deformed theories can be compared asymptotically.
As a measure of the GTE, we employ the concurrence fill~\cite{Xie2021}, which is amenable to analytic treatment.
Because the concurrence fill is not generally an LOCC monotone~\cite{Ge2023,GeWang2024},\footnote{An entanglement measure is called an LOCC monotone if its average value does not increase under any protocol consisting of local operations on the individual subsystems assisted by classical communication between them~\cite{Nielsen1999}.} we also test the same selection rule using the generalized geometric measure (GGM)~\cite{SenDe2010,Das2016,Sadhukhan2017}, thereby avoiding conclusions based solely on the concurrence fill.

We will show that whether $k=1$ is a maximum or a minimum of the concurrence in the corresponding two-particle setup without the spectator depends on the initial helicities, and that neither maximization nor minimization can therefore serve as a single selection principle.
For the $2+1$-particle system, by contrast, we find a common behavior for the family of initial states and color-factor configurations considered here.
Sufficiently close to the forward or backward limit, the concurrence fill and the GGM at $k=1$ are smaller than their values at any fixed $k\neq1$.
Thus, restoring gauge invariance suppresses the generation of GTE more strongly than any fixed deformation in these limits.
Maximizing GTE, however, does not yield a universal finite value of $k$ independent of the color-factor configuration and the initial correlation.

The remainder of this paper is organized as follows.
In Sec.~\ref{sec:amplitudes}, we introduce the kinematics and the gluon and graviton scattering amplitudes with a deformed four-point vertex.
In Secs.~\ref{sec:concurrence} and \ref{sec:tripartite}, we define the concurrence, the concurrence fill, and the GGM used in this work.
In Sec.~\ref{sec:concurrence_and_concurrencefill}, we compare the overall behavior of the concurrence and the concurrence fill and motivate our focus on the forward and backward limits.
In Sec.~\ref{sec:seriesconcurrence}, we analyze the asymptotic expansion of the concurrence in the corresponding two-particle setup without the spectator.
In Sec.~\ref{sec:seriesconcurrencefill}, we analyze the concurrence fill and the GGM in the $2+1$-particle system and show that a fixed-$k$ asymptotic comparison favors $k=1$.
Finally, in Sec.~\ref{sec:discussion}, we summarize our results and present our conclusions.
\section{Deformed amplitudes and kinematics}
\label{sec:amplitudes}
Following Refs.~\cite{Nunez2026Gauge, Nunez2026Universality, Lyu2025,Sannan1986}, we introduce the tree-level gluon and graviton scattering kinematics and the deformation of the four-point vertex used in our analysis.
The explicit tree-level amplitudes are presented in Appendices~\ref{app:amplitudes} and \ref{app:graviton}.

\subsection{Tree-level scattering and kinematics}

Tree-level four-gluon and four-graviton scattering each receive contributions from four Feynman diagrams: the $s$-, $t$-, and $u$-channel diagrams and the four-point contact diagram.
The full scattering amplitude before introducing the deformation can therefore be written as
\begin{align}
 \mathcal{M}
 =\mathcal{M}_{s}+\mathcal{M}_{t}
 +\mathcal{M}_{u}+\mathcal{M}_{4},
 \label{eq:standard-amplitude}
\end{align}
where $\mathcal{M}_{s}$, $\mathcal{M}_{t}$, $\mathcal{M}_{u}$, and $\mathcal{M}_{4}$ are the contributions from the $s$-, $t$-, and $u$-channel diagrams and the four-point contact vertex, respectively.
The kinematics are expressed in terms of the Mandelstam variables $s,t,u$ and the center-of-mass scattering angle $\theta$.

Denoting the four-momenta in the initial state by $p_1,p_2$ and those in the final state by $p_3,p_4$, we have
\begin{align}
 s&=(p_1+p_2)^2,\qquad
 t=(p_1-p_3)^2,\qquad
 u=(p_1-p_4)^2,
 \nonumber\\
 s+t+u&=0,\qquad
 t=-s\sin^2\frac{\theta}{2},\qquad
 u=-s\cos^2\frac{\theta}{2}.
 \label{eq:main-kinematics}
\end{align}
Throughout this paper, we use the mostly-minus metric $\eta_{\mu\nu}=\mathrm{diag}(+,-,-,-)$ for both the gluon and the graviton case, so that $s=E_{\rm cm}^2>0$ and $t,u\leq0$ in the physical region.
For gluon scattering,
\begin{align}
 g^a g^b\longrightarrow g^{a'}g^{b'},
\end{align}
the color dependence is expressed as
\begin{align}
 F_1\equiv \sum_{c=1}^{N^2-1}f^{aa'c}f^{bb'c},\quad
 F_2\equiv \sum_{c=1}^{N^2-1}f^{ab'c}f^{ba'c},\quad
 F_3\equiv \sum_{c=1}^{N^2-1}f^{abc}f^{a'b'c}.
 \label{eq:F123}
\end{align}
Here, $a,b$ are the color indices of the initial-state particles, while $a',b'$ are the color indices of the final-state particles.
The quantities $f^{abc}$ are the structure constants of the $SU(N)$ gauge theory, defined through the commutation relation $[T^a,T^b]=if^{abc}T^c$.
The quantities $F_1,F_2,F_3$ are therefore not color indices themselves, but color factors determined by the external color indices.

\subsection{Deformation of the four-point vertex}

In general, there is no unique way to break gauge invariance.
Following Ref.~\cite{Nunez2026Gauge}, we adopt a minimal tree-level prescription that leaves the free kinetic term unchanged and deforms only the four-point interaction.
We can therefore continue to use the two transverse helicity eigenstates without changing the gluon or graviton degrees of freedom.
This is one particular symmetry-breaking prescription, and whether other deformations obey the same selection rule remains unknown.

In the following, $k\in\mathbb{R}$ denotes a dimensionless deformation parameter.
For QCD, we take the deformed Lagrangian density to be
\begin{align}
 \widetilde{\mathcal{L}}_{\mathrm{QCD}}
 =-\frac{1}{4}F_{\mu\nu}^{a}F^{\mu\nu a}
 -\frac{1}{4}\lambda_{\mathrm{QCD}}\textcolor{black}{\Lambda^{abcd}}
 A_{\mu}^{a}A^{b\mu}A_{\nu}^{c}A^{d\nu},
 \label{eq:deformed-qcd-lagrangian}
\end{align}
where the field strength is given by
\begin{align}
 F_{\mu\nu}^{a}
 =\partial_{\mu}A_{\nu}^{a}
 -\partial_{\nu}A_{\mu}^{a}
 +g f_{bc}^{a}A_{\mu}^{b}A_{\nu}^{c}.
 \label{eq:qcd-field-strength}
\end{align}
\textcolor{black}{Here, $\Lambda^{abcd}$ denotes the color tensor of the additional four-point interaction, and $\Lambda_{\mathrm{QCD}}^{abcd}$ denotes the color tensor of the standard four-gluon vertex of QCD, namely $\Lambda_{\mathrm{QCD}}^{abcd}=\sum_{e}\left(f^{abe}f^{cde}+f^{ace}f^{dbe}+f^{ade}f^{bce}\right)$. We choose $\Lambda^{abcd}=\Lambda_{\mathrm{QCD}}^{abcd}$, so that the additional term has the same color structure as the standard QCD four-gluon vertex, and set}
\begin{align}
 \textcolor{black}{\Lambda^{abcd}=\Lambda_{\mathrm{QCD}}^{abcd}},
 \qquad
 \lambda_{\mathrm{QCD}}\equiv(k-1)g^2.
 \label{eq:qcd-deformation-parameters}
\end{align}
Under this deformation, the Feynman rule for the four-point vertex changes from one proportional to $-ig^2$ to one proportional to $-ikg^2$.
Thus, the gauge-invariant QCD interaction is recovered at $k=1$.

For perturbative quantum gravity, we take the deformed Lagrangian density to be
\begin{align}
 \widetilde{\mathcal{L}}
 =\mathcal{L}_{2}+\mathcal{L}_{3}
 +\mathcal{L}_{4}+\lambda_{\mathrm{GR}}\mathcal{L}_{4}.
 \label{eq:deformed-gr-lagrangian}
\end{align}
\textcolor{black}{Here, $\mathcal{L}_{2}$ is the quadratic kinetic term obtained from the Einstein--Hilbert Lagrangian density, while $\mathcal{L}_{3}$ and $\mathcal{L}_{4}$ are the three- and four-point interactions, respectively.}
We work in the \textcolor{black}{transverse-traceless gauge} for the massless graviton field.
By choosing $\lambda_{\mathrm{GR}}\equiv k-1$, only the four-graviton vertex factor changes at tree level, from one proportional to $-i\kappa^2$ to one proportional to $-ik\kappa^2$.

Accordingly, for both gluons and gravitons, the deformed full amplitude can be written as
\begin{align}
 \mathcal{M}
 =\mathcal{M}_{s}+\mathcal{M}_{t}
 +\mathcal{M}_{u}+k\mathcal{M}_{4}.
 \label{eq:deformation}
\end{align}
Thus, $k=1$ restores gauge invariance in both cases.

\section{Entanglement in $2\to2$ scattering}
\label{sec:concurrence}
In this section, we introduce the concurrence to quantify the entanglement generated by two-particle scattering~\cite{Nunez2026Universality,Nunez2026Gauge}.

In the following, the positive and negative helicities of the external particles are denoted by $+$ and $-$, respectively.
We treat the initial-state particles as incoming external lines and the final-state particles as outgoing external lines, and define the helicity of each particle along its direction of motion.
This convention is used throughout the paper.
For an initial state $\ket{\psi_i}$, the two-particle state after tree-level scattering can be written as
\begin{align}
 \lvert\psi_f\rangle
 \sim{}
 \mathcal{M}_{\psi_i\to ++}\lvert ++\rangle
 +\mathcal{M}_{\psi_i\to +-}\lvert +-\rangle
 +\mathcal{M}_{\psi_i\to -+}\lvert -+\rangle
 +\mathcal{M}_{\psi_i\to --}\lvert --\rangle,
 \label{eq:two-particle-final-state}
\end{align}
where $\mathcal{M}_{\psi_i\to h_Ah_B}$ is the tree-level scattering amplitude for final-state helicities $h_A,h_B\in\{+,-\}$.

The normalized final state in Eq.~\eqref{eq:two-particle-final-state} can be written as
\begin{align}
 \lvert\psi\rangle
 =\alpha\lvert ++\rangle
 +\beta\lvert +-\rangle
 +\gamma\lvert -+\rangle
 +\delta\lvert --\rangle,
 \label{eq:general-two-particle-state}
\end{align}
where the coefficients satisfy $\alpha,\beta,\gamma,\delta\in\mathbb{C}$ and $|\alpha|^2+|\beta|^2+|\gamma|^2+|\delta|^2=1$.
The concurrence of this state is defined by~\cite{Wootters1998}
\begin{align}
 \Delta_{\rm in}=2|\alpha\delta-\beta\gamma|,
 \label{eq:concurrence}
\end{align}
where $\Delta_{\rm in}$ is the concurrence for the initial state $\ket{\psi_i}$ and satisfies $0\leq\Delta_{\rm in}\leq1$.
The cases $\Delta_{\rm in}=0$ and $\Delta_{\rm in}=1$ correspond to a product state and a maximally entangled state, respectively.
\section{$2+1$-particle states and measures of GTE}
\label{sec:tripartite}
In this section, we formulate the scattering of a $2+1$-particle system with an initial bipartite correlation and introduce the concurrence fill and the GGM to characterize the GTE generated after scattering.
The following setup adapts previous studies of correlation transfer using a third particle that does not participate directly in the scattering~\cite{Araujo2019,Fonseca2022,Blasone2024,Cao2026} to the gluon and graviton scattering processes considered here.

\subsection{Setup for $2+1$-particle scattering}

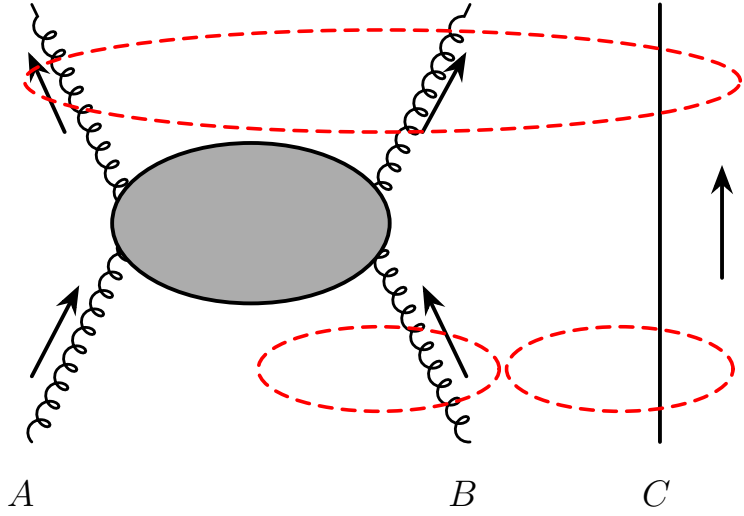
\begin{figure}[!t]
\centering
\resizebox{0.65\textwidth}{!}{%
\begin{tikzpicture}[
    line cap=round,
gluon/.style={
    draw=black,
    line width=1.1pt,
    decorate,
    decoration={
        coil,
        aspect=0.8,
        segment length=8pt,
        amplitude=3.5pt
    }
},
    spectator/.style={
        black,
        line width=1.5pt
    },
    spin/.style={
        black,
        line width=1.5pt,
        -{Stealth[length=4mm,width=3mm]}
    },
    redloop/.style={
        red,
        line width=1.4pt,
        dash pattern=on 5pt off 4pt
    },
    label/.style={
        font=\Large
    }
]
\draw[gluon] (-3.0,-3.0) -- (-1.55,0.0);
\draw[gluon] (-1.55,0.0) -- (-3.0,3.0);
\draw[spin] (-3.0,-2.1) -- (-2.35,-0.85);
\draw[spin] (-2.55,1.25) -- (-3.05,2.35);
\draw[gluon] (3.0,-3.0) -- (1.55,0.0);
\draw[gluon] (1.55,0.0) -- (3.0,3.0);
\draw[spin] (2.95,-2.1) -- (2.35,-0.85);
\draw[spin] (2.35,1.25) -- (2.95,2.35);
\draw[spectator] (5.6,-3.0) -- (5.6,3.0);
\draw[spin] (6.45,-0.75) -- (6.45,0.80);
\draw[
    fill=gray!65,
    draw=black,
    line width=1.5pt
]
(0,0) ellipse [x radius=1.9, y radius=1.1];
\draw[redloop]
    (1.8,1.95)
    ellipse [x radius=4.9, y radius=0.70];
\draw[redloop]
    (1.75,-2.00)
    ellipse [x radius=1.65, y radius=0.58];
\draw[redloop]
    (5.05,-2.00)
    ellipse [x radius=1.55, y radius=0.58];
\node[label] at (-3.15,-3.70) {$A$};
\node[label] at ( 2.90,-3.70) {$B$};
\node[label] at ( 5.55,-3.70) {$C$};
\end{tikzpicture}%
}
\caption{
Schematic representation of the $2+1$-particle system.
Particles $B$ and $C$ are initially entangled, and scattering between
$A$ and $B$ can generate a tripartite correlation among $A$, $B$, and $C$.
Here, particle $C$ is assumed to remain spatially separated from $A$ and $B$ throughout the process.}
\label{fig:diagram}
\end{figure}

As illustrated in Fig.~\ref{fig:diagram}, particles $B$ and $C$ are entangled before the scattering, and the subsequent scattering of particles $A$ and $B$ can redistribute this bipartite correlation among all three particles.
We take particle $A$ to have positive helicity and take the initial $BC$ state to be
\begin{align}
 \ket{\Psi_{\rm in}}
 =\ket{+}_A\otimes
 \left(\cos\eta\ket{++}_{BC}+e^{i\zeta}\sin\eta\ket{--}_{BC}\right),
 \label{eq:initial}
\end{align}
where \textcolor{black}{$0<\eta<\pi/2$} is a real parameter controlling the relative weights of the two components of the initial $BC$ state.
Because the value of $\zeta$ does not affect the entanglement measures considered below, we set $\zeta=0$.
We restrict the analysis to this subclass of initial states, which contains no additional single-particle helicity superpositions.

Denoting the transition operator for the scattering by $\widehat{\mathcal{T}}_{AB}$, the operator acting on the tripartite system can be written as
\begin{align}
 \widehat{\mathcal{T}}_{ABC}
 =\widehat{\mathcal{T}}_{AB}\otimes\mathbb{I}_C,
 \label{eq:three-particle-scattering-operator}
\end{align}
where $\mathbb{I}_C$ is the identity operator on the Hilbert space of particle $C$.
The indices of the amplitude are ordered as
\begin{align}
 \mathcal{M}_{\lambda_A\lambda_B\to\lambda_A'\lambda_B'}
 \equiv
 \bra{\lambda_A'\lambda_B'}_{AB}
 \widehat{\mathcal{T}}_{AB}
 \ket{\lambda_A\lambda_B}_{AB},
 \label{eq:helicity-amplitude-convention}
\end{align}
where $\lambda_A,\lambda_B$ and $\lambda_A',\lambda_B'$ are the initial- and final-state helicities, respectively.
We suppress the dependence of the $\mathcal{M}$ matrix on the momenta, scattering angle, and \textcolor{black}{external color indices}.
The explicit expressions for the amplitudes are given in Appendices~\ref{app:amplitudes} and \ref{app:graviton}.

Applying Eq.~\eqref{eq:three-particle-scattering-operator} to Eq.~\eqref{eq:initial}, the unnormalized final state after scattering can be written in terms of the tree-level amplitudes as
\begin{align}
 \ket{\widetilde\Psi_{\rm out}}
 &=\cos\eta\,\mathcal{M}_{++\to++}\ket{+++}
 +\cos\eta\,\mathcal{M}_{\rm cross}\ket{+-+}\nonumber\\
 &+\cos\eta\,\mathcal{M}_{\rm cross}\ket{-++}
 +\cos\eta\,\mathcal{M}_{++\to--}\ket{--+}\nonumber\\
 &+\sin\eta\,\mathcal{M}_{\rm cross}\ket{++-}
 +\sin\eta\,\mathcal{M}_{+-\to+-}\ket{+--}\nonumber\\
 &+\sin\eta\,\mathcal{M}_{+-\to-+}\ket{-+-}
 +\sin\eta\,\mathcal{M}_{\rm cross}\ket{---}.
 \label{eq:outstate}
\end{align}
Here, $\mathcal{M}_{\rm cross}$ denotes the single-helicity-flip amplitudes, which are all equal by symmetry,
\begin{align}
 \mathcal{M}_{\rm cross}
 \equiv
 \mathcal{M}_{\substack{++\\--}\to\substack{+-\\-+}}
 =
 \mathcal{M}_{\substack{+-\\-+}\to\substack{++\\--}},
 \label{eq:Mcross-definition-main}
\end{align}
as shown explicitly in Appendices~\ref{app:amplitudes} and \ref{app:graviton}.

We therefore define the normalized final state post-selected on the specified final-state momenta and, for gluon scattering, the specified external-color configuration, together with its density matrix, by
\begin{align}
 \ket{\Psi_{\rm out}}
 &=\frac{\ket{\widetilde\Psi_{\rm out}}}
 {\textcolor{black}{\sqrt{
 \braket{\widetilde\Psi_{\rm out}}
 }}},
 \nonumber\\
 \rho_{ABC}
 &=\ket{\Psi_{\rm out}}\bra{\Psi_{\rm out}}.
 \label{eq:normalized-out-state}
\end{align}
We exclude points at which $\braket{\widetilde\Psi_{\rm out}}=0$, because the normalized final state is undefined there.

\subsection{Concurrence fill}

The concurrence fill provides a geometric indicator of genuine tripartite entanglement through the area of the triangle formed by the three squared one-versus-rest concurrences~\cite{Xie2021,Cao2026}.

To extract information about each subsystem from the final state in Eq.~\eqref{eq:normalized-out-state}, we introduce the partial trace.
Let the Hilbert space of a composite system be $\mathcal{H}=\mathcal{H}_{X}\otimes\mathcal{H}_{Y}$, where $Y$ denotes the subsystem complementary to $X$.
For a complete orthonormal basis $\{\ket{x_n}_{X}\}$ of $\mathcal{H}_{X}$, the partial trace over $X$ is defined for an arbitrary operator $\mathcal{O}$ by
\begin{align}
 \Tr_{X}\mathcal{O}
 \equiv
 \sum_n
 \left(\bra{x_n}_{X}\otimes\mathbb{I}_{Y}\right)
 \mathcal{O}
 \left(\ket{x_n}_{X}\otimes\mathbb{I}_{Y}\right),
 \label{eq:partial-trace-definition}
\end{align}
where $\mathbb{I}_{Y}$ is the identity operator on $\mathcal{H}_{Y}$.
This definition is independent of the choice of basis $\{\ket{x_n}_{X}\}$, and $\Tr_X\mathcal{O}$ is an operator on $\mathcal{H}_{Y}$.
Thus, for a density matrix $\rho_{XY}$ of the composite system, the reduced density matrix obtained by tracing out the degrees of freedom of subsystem $X$ is
\begin{align}
 \rho_{Y}
 =\Tr_X\rho_{XY}.
 \label{eq:reduced-density-matrix-definition}
\end{align}

For the final state in Eq.~\eqref{eq:normalized-out-state}, the three single-qubit reduced density matrices are defined by
\begin{align}
 \rho_A&=\Tr_{BC}\rho_{ABC},\quad
 \rho_B=\Tr_{AC}\rho_{ABC}, \quad
 \rho_C=\Tr_{AB}\rho_{ABC}.
\end{align}
For a pure three-qubit state, the one-versus-rest concurrence between subsystem $X$ and the remaining two-qubit system $Y$ is
\begin{align}
 C_{X|Y}
 =2\sqrt{\det\rho_X}
 =\sqrt{2\left[1-\Tr\!\left(\rho_X^2\right)\right]}.
 \label{eq:one-versus-rest-concurrence}
\end{align}
We then define the three squared concurrences as
\begin{align}
 a=C_{A|BC}^2=4\det\rho_A,\quad
 b=C_{B|AC}^2=4\det\rho_B,\quad
 c=C_{C|AB}^2=4\det\rho_C,
 \label{eq:abc}
\end{align}
and regard $a,b,c$ as the three sides of the concurrence triangle~\cite{Coffman2000}.
The semiperimeter of this triangle is
\begin{align}
 Q=\frac{a+b+c}{2},
 \label{eq:concurrence-semiperimeter}
\end{align}
and Heron's formula gives its area as
\begin{align}
 \mathcal{A}_{\rm C}
 =\sqrt{Q(Q-a)(Q-b)(Q-c)}.
 \label{eq:concurrence-triangle-area}
\end{align}
The concurrence fill is defined as the normalized area
\begin{align}
 \CF
 =\left[\frac{16}{3}Q(Q-a)(Q-b)(Q-c)\right]^{1/4}.
 \label{eq:CF}
\end{align}
The factor $16/3$ ensures that $0\leq\CF\leq1$.
For example, the normalized version of $\ket{+++}+\ket{---}$ gives $a=b=c=1$ and hence $\CF=1$.
By contrast, for the normalized version of a biseparable state such as $\ket{+++}+\ket{+--}$, one has $C_{A|BC}=0$ and $C_{B|AC}=C_{C|AB}$, so that $\CF=0$.
Geometrically, this occurs because one side of the triangle has zero length.
Accordingly, the initial state in Eq.~\eqref{eq:initial} has $\CF=0$.

The concurrence fill is symmetric under permutations of $A,B,C$ and simultaneously captures both the amount of entanglement across each of the three one-versus-rest bipartitions and the balance among them as a single geometric quantity.
Moreover, because $\CF^4$ is free of fourth roots and is differentiable throughout the domain in which the normalized final state is defined, it is convenient for analytic treatment of the forward expansion.

We use the concurrence fill as a geometric indicator of GTE; however, it is not generally an LOCC monotone~\cite{Ge2023, Xie2021,GeWang2024}.
We therefore compare it with an independent quantitative measure of GTE.

\subsection{Generalized geometric measure}

The GGM quantifies the degree of GTE in a pure state of an arbitrary $n$-particle system~\cite{SenDe2010,Das2016,Sadhukhan2017,Cao2026}.
For a pure three-qubit state $\ket{\Psi}$, the GGM is defined as
\begin{align}
 \mathcal{E}_{\mathrm{GGM}}(\ket{\Psi})
 &=
 1-\max_{X=A,B,C}\lambda_{\max}(\rho_X)
 \nonumber\\
 &=
 \min_{X=A,B,C}g_X,
 \label{eq:GGM-definition}
\end{align}
where $\lambda_{\max}(\rho_X)$ is the largest eigenvalue of the reduced density matrix $\rho_X$.
In terms of the squared concurrences $a,b,c$ defined in Eq.~\eqref{eq:abc}, the candidates associated with the three bipartitions are
\begin{align}
 g_A&\equiv\frac{1-\sqrt{1-a}}{2},&
 g_B&\equiv\frac{1-\sqrt{1-b}}{2},&
 g_C&\equiv\frac{1-\sqrt{1-c}}{2}.
 \label{eq:GGM-branches}
\end{align}
The GGM is therefore the smallest of the three candidate functions $g_A,g_B,g_C$, corresponding respectively to the bipartitions $A|BC$, $B|AC$, and $C|AB$.

Different measures capture different aspects of GTE.
Here, we use the GGM as an independent quantitative check of the result obtained with the concurrence fill.

\section{Numerical comparison of concurrence and concurrence fill}
\label{sec:concurrence_and_concurrencefill}
In this section, we numerically compare the concurrence and the concurrence fill.
The point $\theta=\pi/2$, which is special in the concurrence analysis, is no longer a special point for the concurrence fill; instead, the forward and backward regimes exhibit the clearest structure.
Because the purpose of this section is solely to motivate our focus on the vicinity of forward and backward scattering, we do not analyze the GGM here.
The gluon and graviton scattering amplitudes and the procedures used to evaluate the two measures are given in Appendices~\ref{app:amplitudes} and \ref{app:graviton}.

\subsection{Behavior of the concurrence at $\theta=\pi/2$ and $\pi/6$}

\begin{figure}[!t]
  \centering
  \begin{subfigure}{0.48\columnwidth}
    \centering
    \IfFileExists{
      AverageConcurrence_RL_theta_Pi_2_with_graviton.pdf
    }{%
      \includegraphics[
        width=\linewidth,
        keepaspectratio
      ]{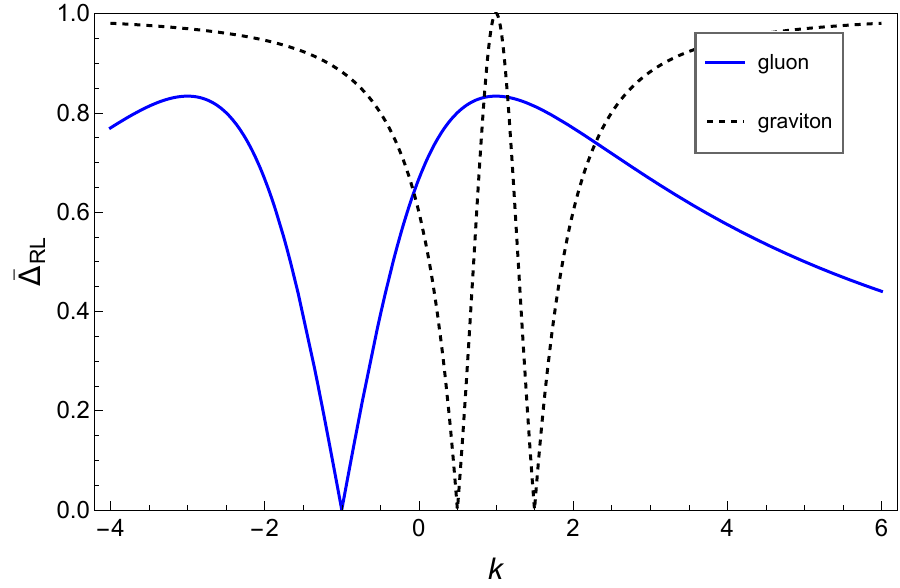}
    }{%
      \fbox{%
        \parbox[c][32mm][c]{0.88\linewidth}{%
          \centering
          Place the figure file in the same directory
        }%
      }%
    }
    \caption{\textcolor{black}{Initial helicities $+-$}}
    \label{fig:concurrenceRL-pi2}
  \end{subfigure}
  \hfill
  \begin{subfigure}{0.48\columnwidth}
    \centering
    \IfFileExists{
      AverageConcurrence_RR_theta_Pi_2_with_graviton.pdf
    }{%
      \includegraphics[
        width=\linewidth,
        keepaspectratio
      ]{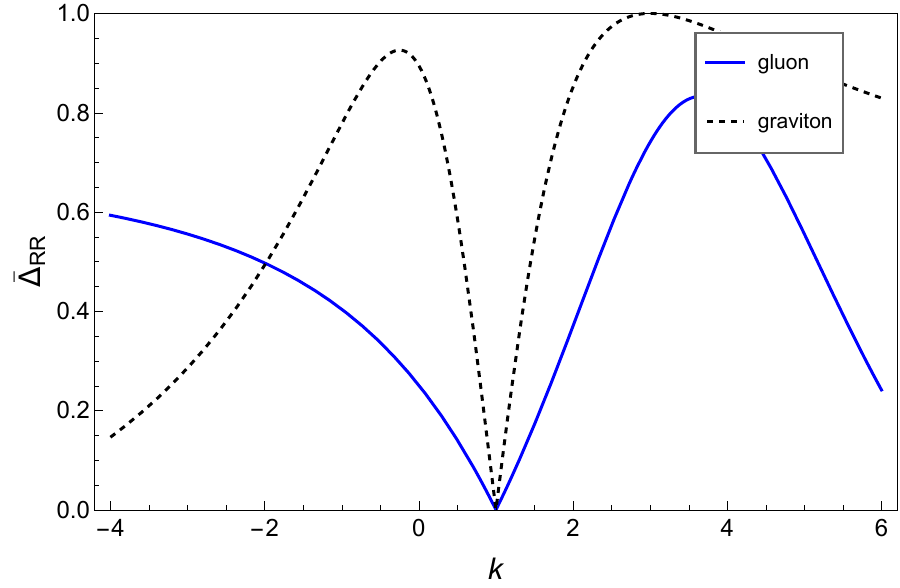}
    }{%
      \fbox{%
        \parbox[c][32mm][c]{0.88\linewidth}{%
          \centering
          Place the figure file in the same directory
        }%
      }%
    }
    \caption{\textcolor{black}{Initial helicities $++$}}
    \label{fig:concurrenceRR-pi2}
  \end{subfigure}
  \caption{\textcolor{black}{Configuration-averaged concurrence as a function of $k$ at the scattering angle $\theta=\pi/2$.
  The results of Ref.~\cite{Nunez2026Gauge} are reproduced using the scattering amplitudes given therein.}}
  \label{fig:concurrence-pi2}
\end{figure}

\begin{figure}[!t]
  \centering
  \begin{subfigure}{0.48\columnwidth}
    \centering
    \IfFileExists{
      AverageConcurrence_RL_theta_Pi_6_with_graviton.pdf
    }{%
      \includegraphics[
        width=\linewidth,
        keepaspectratio
      ]{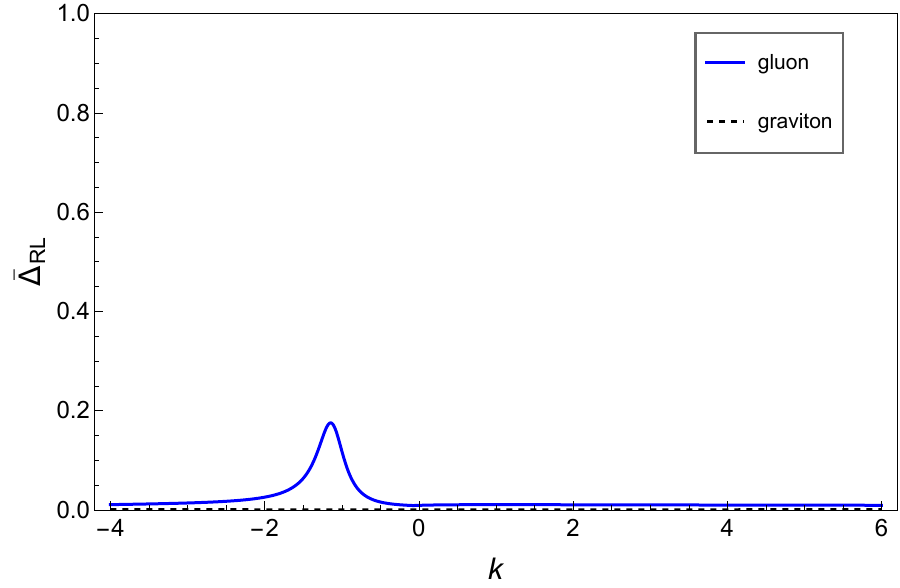}
    }{%
      \fbox{%
        \parbox[c][32mm][c]{0.88\linewidth}{%
          \centering
          Place the figure file in the same directory
        }%
      }%
    }
    \caption{\textcolor{black}{Initial helicities $+-$}}
    \label{fig:concurrenceRL-pi6}
  \end{subfigure}
  \hfill
  \begin{subfigure}{0.48\columnwidth}
    \centering
    \IfFileExists{
      AverageConcurrence_RR_theta_Pi_6_with_graviton.pdf
    }{%
      \includegraphics[
        width=\linewidth,
        keepaspectratio
      ]{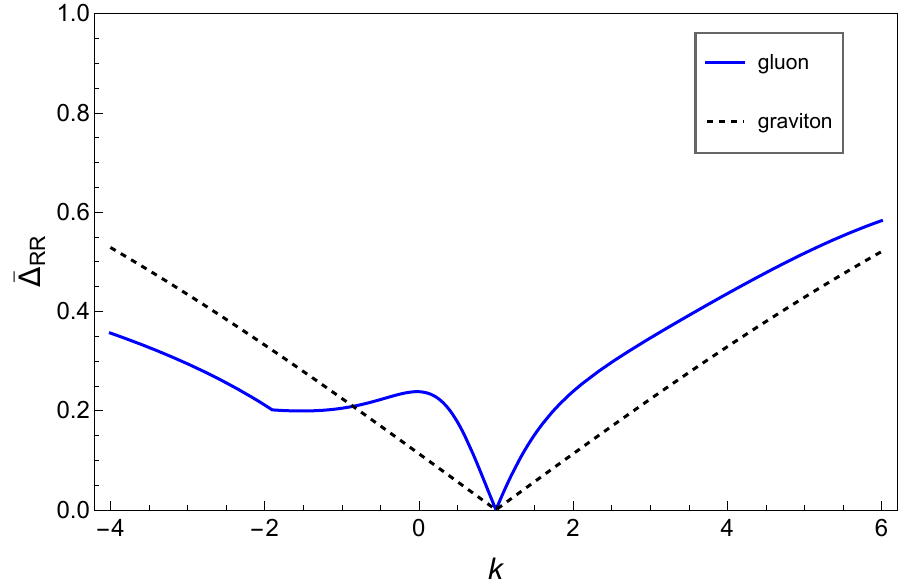}
    }{%
      \fbox{%
        \parbox[c][32mm][c]{0.88\linewidth}{%
          \centering
          Place the figure file in the same directory
        }%
      }%
    }
    \caption{\textcolor{black}{Initial helicities $++$}}
    \label{fig:concurrenceRR-pi6}
  \end{subfigure}
  \caption{\textcolor{black}{Configuration-averaged concurrence as a function of $k$ at the scattering angle $\theta=\pi/6$.}}
  \label{fig:concurrence-pi6}
\end{figure}

Figure~\ref{fig:concurrence-pi2} reproduces the behavior at $\theta=\pi/2$ reported in Refs.~\cite{Nunez2026Universality,Nunez2026Gauge}, while Fig.~\ref{fig:concurrence-pi6} shows the corresponding behavior at the representative nonspecial angle $\theta=\pi/6$, which is not presented in those references.
To visualize trends common to several representative external-color configurations, we introduce the configuration-averaged concurrence
\begin{align}
    \bar{\Delta}_{\rm in} \equiv \frac{1}{N_{\rm cfg}}\sum_{\alpha\in\mathcal{C}}\Delta_{{\rm in},\alpha},
 \label{eq:avgconcrrence}
\end{align}
where we average over the following set of color-factor configurations~\cite{Nunez2026Gauge}:
\begin{revision}
\begin{align}
 \mathcal{C}
 =\bigl\{&(0,1,-1),(1,0,1),(1,1,0),
(2,1,1),(1,2,-1),(1,-1,2)\bigr\}.
 \label{eq:configuration-set}
\end{align}
\end{revision}
Each entry of Eq.~\eqref{eq:configuration-set} is an ordered triple $(F_1,F_2,F_3)$ of the color factors defined in Eq.~\eqref{eq:F123}.
\textcolor{black}{Here,} $\alpha$ labels a color-factor configuration, and $N_{\rm cfg}=|\mathcal{C}|$ is the number of configurations.
We use this quantity only in this section to illustrate the overall behavior of the concurrence.

\begin{revision}
For visualization, we use an unweighted arithmetic mean over the color-factor classes; this is not a physical color average over initial colors or a sum over final-state colors.
The principal results below are therefore based on separate analyses of the individual color-factor configurations rather than on this configuration average.
\end{revision}

Figure~\ref{fig:concurrence-pi2} confirms that, at $\theta=\pi/2$, $k=1$ is an extremum: it is a maximum or a minimum depending on the initial helicities.
This observation does not establish whether maximization or minimization of entanglement is the operative principle.
At a representative nonspecial angle such as $\theta=\pi/6$, Fig.~\ref{fig:concurrence-pi6} shows that $k=1$ does not exhibit comparably simple extremal behavior.

\subsection{Behavior of the concurrence fill at $\theta=\pi/2$ and $\pi/6$}

\begin{figure}[!t]
  \centering
  \begin{subfigure}{0.48\columnwidth}
    \centering
    \IfFileExists{
      AverageCF_theta_Pi_2_eta_Pi_4_with_graviton.pdf
    }{%
      \includegraphics[
        width=\linewidth,
        keepaspectratio
      ]{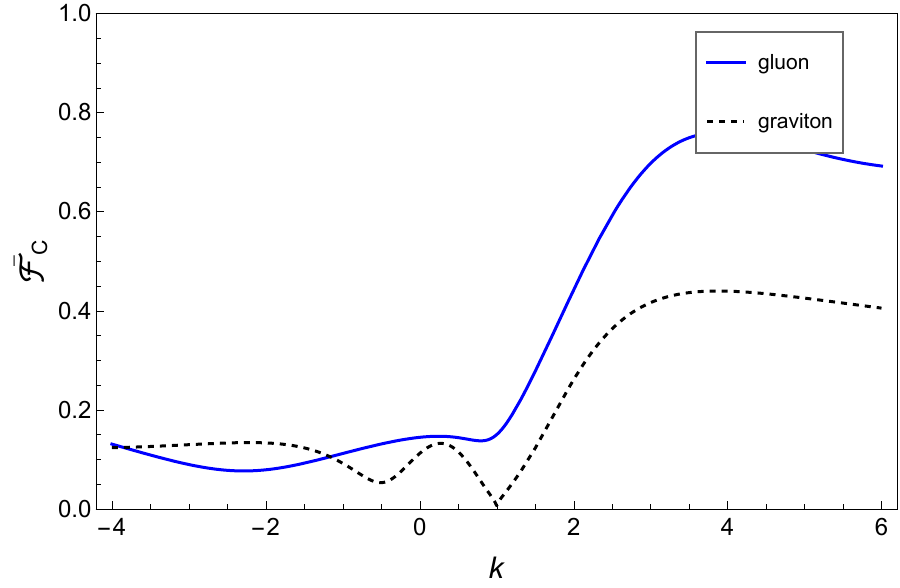}
    }{%
      \fbox{%
        \parbox[c][32mm][c]{0.88\linewidth}{%
          \centering
          Place the figure file in the same directory
        }%
      }%
    }
    \caption{$\theta=\pi/2$}
    \label{fig:gluongraviton-pi2}
  \end{subfigure}
  \hfill
  \begin{subfigure}{0.48\columnwidth}
    \centering
    \IfFileExists{
      AverageCF_theta_Pi_6_eta_Pi_4_with_graviton.pdf
    }{%
      \includegraphics[
        width=\linewidth,
        keepaspectratio
      ]{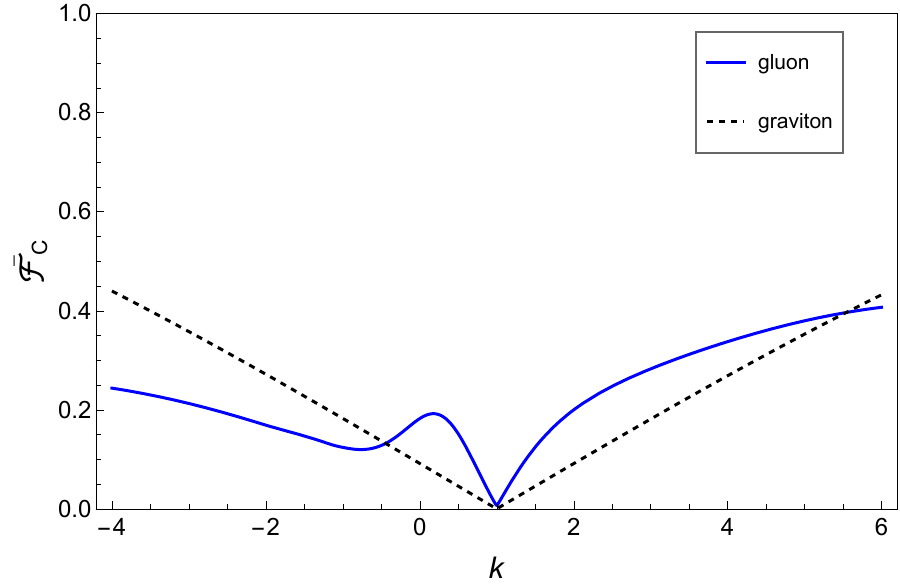}
    }{%
      \fbox{%
        \parbox[c][32mm][c]{0.88\linewidth}{%
          \centering
          Place the figure file in the same directory
        }%
      }%
    }
    \caption{$\theta=\pi/6$}
    \label{fig:gluongraviton-pi6}
  \end{subfigure}
  \caption{\textcolor{black}{Configuration-averaged concurrence fill as a function of $k$ for the initial-state parameter $\eta=\pi/4$.
  The color-factor configurations in Eq.~\eqref{eq:configuration-set} are included in the average.
  Panels (a) and (b) show the results for the scattering angles $\theta=\pi/2$ and $\theta=\pi/6$, respectively.}}
  \label{fig:gluongraviton}
\end{figure}

We now analyze the concurrence fill.
As in the concurrence analysis, we introduce the configuration-averaged concurrence fill
\begin{align}
 \overline{\CF}
 \equiv\frac{1}{N_{\rm cfg}}\sum_{\alpha\in\mathcal{C}}\mathcal{F}_{{\rm C},\alpha}.
 \label{eq:avgCF}
\end{align}
Like Eq.~\eqref{eq:avgconcrrence}, we use this quantity only in the present section.
To examine the capacity of scattering to redistribute correlations, we set the initial-state parameter to $\eta=\pi/4$, corresponding to a maximally entangled initial $BC$ state, and analyze $\theta=\pi/2$ and $\theta=\pi/6$ for comparison with the concurrence.

We find that, for gluon scattering at $\theta=\pi/2$, the concurrence fill has no special feature at $k=1$, as shown in Fig.~\ref{fig:gluongraviton}.
For graviton scattering, we also find that the location of the minimum is slightly displaced from $k=1$.
At $\theta=\pi/6$, by contrast, we find that the minimum lies much closer to $k=1$, and analogous behavior is observed as the backward limit $\theta\to\pi$ is approached.
As the scattering angle decreases, the location of the minimum approaches $k=1$ but need not coincide with it at finite angle.
These observations lead us to expect the following ordering: once a comparison point $k=k_{*}\neq1$ is fixed in advance, the concurrence fill at $k=1$ is smaller than its value at $k=k_{*}$ for all sufficiently small scattering angles.
Specifically, for any fixed $k_{*}\neq1$, we investigate whether there exists a $\theta_0(k_{*},\eta;\alpha)>0$ such that
\begin{align}
 \CF(k=1,\theta,\eta;\alpha)
 <
 \CF(k=k_{*},\theta,\eta;\alpha),
 \qquad
 0<\theta<\theta_0(k_{*},\eta;\alpha).
 \label{eq:kpredictionCF}
\end{align}

In what follows, we refer to this procedure as a \emph{fixed-$k$ asymptotic comparison}: the comparison point $k_{*}$, the initial correlation $\eta$, and the color-factor configuration $\alpha$ are all fixed before the limit $\theta\to0$ is taken.
Here, $\theta_0(k_{*},\eta;\alpha)$ may differ for each comparison point; the existence of a common $\theta_0$ for all $k_{*}$ is not required.
Equation~\eqref{eq:kpredictionCF} therefore does not imply that $k=1$ is the exact global minimizer with respect to $k$ at every finite $\theta$.
In Sec.~\ref{sec:seriesconcurrencefill}, we verify Eq.~\eqref{eq:kpredictionCF} analytically.

\section{Series structure of the concurrence in gluon scattering}
\label{sec:seriesconcurrence}
For comparison with the results obtained with the concurrence fill, we examine the small-angle expansion of the concurrence.
For gluon scattering, $k=1$ minimizes the concurrence for the initial state $\ket{++}$ but locally maximizes it for the initial state $\ket{+-}$.
\textcolor{black}{As a concrete effect of gauge invariance on forward scattering, we also show that the tree-level equal-helicity pair-flip amplitude $\ket{++}\leftrightarrow\ket{--}$ vanishes at $k=1$.}

\subsection{Initial state $\ket{++}$}
\label{subsec:seriesconcurrenceRR}

The unnormalized final state can generally be written as
\begin{align}
 \ket{\widetilde\Phi_{\rm out}}_{++}
 =\mathcal{M}_{++\to++}\ket{++}
 +\mathcal{M}_{\rm cross}\ket{+-}+\mathcal{M}_{\rm cross}\ket{-+}
 +\mathcal{M}_{++\to--}\ket{--}.
 \label{eq:RRoutstateconcurrence}
\end{align}
Upon setting $k=1$, however, Appendix~\ref{app:amplitudes} gives $\mathcal{M}_{\rm cross}=\mathcal{M}_{++\to--}=0$\footnote{This is a tree-level statement. At one loop, the amplitudes $\mathcal{M}_{++\to--}$ and $\mathcal{M}_{\rm cross}$ are known to be nonvanishing even at $k=1$. Such loop-induced contributions are suppressed relative to the tree-level amplitudes and do not affect the results obtained in this work.}, so that
\begin{align}
 \ket{\widetilde\Phi_{\rm out}(k=1)}_{++}
 =\mathcal{M}_{++\to++}\ket{++}.
 \label{eq:k1RRoutstateconcurrence}
\end{align}
Because this is manifestly a product state, the concurrence vanishes at $k=1$ for every scattering angle,
\begin{align}
 \Delta_{++}^{(1)}=0,
 \label{eq:DeltaRR-k1-zero}
\end{align}
where $\Delta_{++}^{(1)}$ is the concurrence at $k=1$ for the initial state $\ket{++}$.
Consequently, the concurrence is minimized at $k=1$ for every scattering angle, including the small-angle regime.
We do not analyze extrema away from $k=1$.

\begin{revision}
To express the forward expansion compactly, we define
\begin{align}
 q\equiv\sin^2\frac{\theta}{2}.
 \label{eq:q-definition}
\end{align}
\end{revision}
For the special color-factor configuration $(F_1,F_2,F_3)=(0,1,-1)$, the forward expansion of the concurrence is
\begin{align}
    \Delta_{++} = \frac{4|k^2-1|}{(k+1)^2+4(k-1)^2}q^0 + \mathcal{O}(q),
    \label{eq:concurrenceRR0}
\end{align}
and hence a nonzero amount of \textcolor{black}{entanglement} is generated in the forward limit for $k\neq\pm1$.
In this color-factor configuration the $t$-channel contribution is absent, so that a finite amplitude survives in the forward limit.
The forward entanglement in Eq.~\eqref{eq:concurrenceRR0} is then generated by the equal-helicity pair-flip amplitude, which vanishes in the gauge-invariant theory at $k=1$.
Such color-factor configurations also occur in massless $SU(2)$ \textcolor{black}{Yang--Mills} theory, providing a concrete example of gauge invariance affecting forward scattering.
Appendix~\ref{app:helicityflip} verifies this statement explicitly.

Thus, the minimum for the initial state $\ket{++}$ is consistent with the concurrence-fill inequality in Eq.~\eqref{eq:kpredictionCF}.

\begin{revision}
The forward- and backward-scattering amplitudes are related by the transformation $(F_1,F_2,F_3)\rightarrow(F_2,F_1,-F_3)$.
For example, the color-factor configuration $(0,1,-1)$ in forward scattering corresponds to $(1,0,1)$ in backward scattering.
The concurrence and the concurrence fill are symmetric under interchange of the final-state particles $A$ and $B$.
Their series structures and the fixed-$k$ asymptotic comparison therefore carry over to the corresponding backward-scattering process.
This correspondence is demonstrated in Appendix~\ref{app:forwardandback}.
\end{revision}

\subsection{Initial state $\ket{+-}$}

\subsubsection{$F_1\neq 0$}

For the initial state $\ket{+-}$, the situation differs from the preceding case.
The unnormalized final state is
\begin{align}
 \ket{\widetilde\Phi_{\rm out}}_{+-}
 &=\mathcal{M}_{\rm cross}\ket{++}
 +\mathcal{M}_{+-\to+-}\ket{+-}
 +\mathcal{M}_{+-\to-+}\ket{-+}
 +\mathcal{M}_{\rm cross}\ket{--},
 \label{eq:RLoutstateconcurrence}
\end{align}
and the result at $k=1$ depends on the scattering angle.
We must therefore compute the small-$q$ expansion of the concurrence.
Expanding the concurrence in the variable $q$ defined in Eq.~\eqref{eq:q-definition}, we obtain
\begin{align}
    \Delta_{+-} = 2q^2 + 4q^3+ \left[ 6-\frac{(F_1+F_2)^2}{2F^2_1}(k-1)^2 \right]q^4 + \mathcal{O}(q^5).
    \label{eq:RLseriesconcurrence}
\end{align}
Thus, for $F_1+F_2\neq0$, $k=1$ is indeed a local maximum.
When $F_1+F_2=0$, $\Delta_{+-}$ is independent of $k$, as is evident from Eq.~\eqref{eq:Mdeformedappendix}.
Color configurations satisfying this condition exist in gluon scattering.

\subsubsection{$F_1 = 0, F_2\neq0$}

For $F_1=0$, the analogous series expansion gives
\begin{align}
    \Delta_{+-} &= \frac{8|k|}{(k+1)^2}q^2 + \mathcal{O}(q^3) \qquad (k\neq-1),\notag\\
    \Delta_{+-} &= \frac23 - \frac23q^2 + \mathcal{O}(q^3) \qquad (k=-1).
    \label{eq:RLseriesconcurrence2}
\end{align}
In this expression, $k=1$ is a local maximum but not the global maximum. The result in Eq.~\eqref{eq:RLseriesconcurrence2} is independent of $F_2$.
When $F_1=F_2=0$, all relevant amplitudes vanish, so the normalized final state, and hence $\Delta_{+-}$, is undefined.

Accordingly, for the initial state $\ket{+-}$, $k=1$ is a local maximum for color-factor configurations with nontrivial $k$ dependence.
For $F_1+F_2=0$, by contrast, the concurrence is independent of $k$.
This behavior contrasts with that of the initial state $\ket{++}$.

\section{Series structure of the concurrence fill and the GGM}
\label{sec:seriesconcurrencefill}
In this section, we test the fixed-$k$ asymptotic comparison defined by
Eq.~\eqref{eq:kpredictionCF}.
Because $\CF$ and $\CF^4$ have the same ordering, we expand $\CF^4$
in a series in the scattering angle $\theta$.
We then perform an analogous analysis of the GGM to determine whether
the resulting selection rule is reproduced by an independent measure of GTE.

\subsection{Series structure of the concurrence fill}
The expansion results for the concurrence fill are summarized in
Table~\ref{tab:leadingorders}.
To classify the results, we introduce the following auxiliary functions:

\begin{align}
 \mathcal{P}_1(k,\eta)
 &\equiv
 \frac{1}{3}(k-1)^4\cos^{10}\eta\sin^6\eta,
 \label{eq:Pdef}\\
 \mathcal{G}_1(k,\eta)
 &\equiv
 -\frac{32768}{3}
 (k-1)^4(k+1)^{10}\cos^{10}\eta\sin^4\eta
 \frac{
 -9+14k-9k^2+(k+1)^2\cos2\eta
 }{
 \left[
 3-2k+3k^2+2(k-1)^2\cos2\eta
 \right]^8
 },
 \label{eq:G1def}\\
 \mathcal{H}_1(k,\eta)
 &\equiv
 \frac{\sin^6(2\eta)}{196608}
 \left[4+9(k-1)^2\right]
 \left[
 4(1-\cos2\eta)^2
 +9(k-1)^2(1+\cos2\eta)^2
 \right].
 \label{eq:H1def}
\end{align}

\begin{table}[!t]
\begin{minipage}{0.98\textwidth}
 \centering
 \captionof{table}{
 Leading terms for generic $k$ in the forward expansion of $\CF^4$
 for gluon and graviton scattering.
 $(F_1,F_2,F_3)$ denotes the color-factor configuration.
 Neither a common nonzero rescaling of all $F_i$ nor an overall sign
 change affects the normalized $\CF$.
 The correspondence with backward scattering is shown in
 Appendix~\ref{app:forwardandback}.
 }
 \label{tab:leadingorders}
 \normalsize
 \renewcommand{\arraystretch}{1.2}
 \setlength{\tabcolsep}{7pt}
 \begin{adjustbox}{max width=\linewidth}
 \begin{tabular}{cc}
  \toprule
  $(F_1,F_2,F_3)$ or graviton
  & leading term \\
  \midrule
  $(0,1,-1)$
  & $\mathcal{G}_1(k,\eta)\theta^0$
  \\[2mm]
  $(1,0,1),\ (1,1,0)$
  & $\mathcal{P}_1(k,\eta)\theta^8$
  \\[2mm]
  $(1,2,-1),\ (1,-1,2)$
  & $81\mathcal{P}_1(k,\eta)\theta^8$
  \\[2mm]
  $(2,1,1)$
  & $\mathcal{H}_1(k,\eta)\theta^{16}$
  \\[2mm]
  graviton
  & $4096\mathcal{P}_1(k,\eta)\theta^{16}$
  \\
  \bottomrule
 \end{tabular}
 \end{adjustbox}
\end{minipage}
\end{table}

The results in Table~\ref{tab:leadingorders} fall into three classes.
First, for color-factor configurations whose leading terms are
proportional to $\mathcal{P}_1(k,\eta)$ and for graviton scattering,
the leading term contains an explicit factor $(k-1)^4$ and therefore
vanishes at $k=1$.
Equation~\eqref{eq:kpredictionCF} consequently holds for this class.

For color-factor configurations of the $\mathcal{H}_1$ type,
the leading term does not generally vanish at $k=1$.
However, both square brackets in Eq.~\eqref{eq:H1def} contain
a non-negative increment proportional to $(k-1)^2$.
Thus, the leading coefficient at any fixed $k_{*}\neq1$ is larger than
the coefficient at $k=1$.
The asymptotic comparison in Eq.~\eqref{eq:kpredictionCF} therefore
also holds for this class.

For color-factor configurations of the $\mathcal{G}_1$ type,
the leading term in Eq.~\eqref{eq:G1def} is positive for a fixed
$k_{*}\neq\pm1$, whereas it vanishes at $k=\pm1$.
The only exceptional branches remaining in the asymptotic comparison
with a fixed deformed theory are therefore $k=1$ and $k=-1$,
and these two values must be compared at the first order at which
they differ.
As shown in Appendix~\ref{app:G1structure}, their difference first
appears at order $\theta^{16}$, and for sufficiently small $\theta$,
\begin{align}
 \CF(k=1,\theta,\eta)
 <
 \CF(k=-1,\theta,\eta).
\end{align}
Thus, Eq.~\eqref{eq:kpredictionCF} also holds for this configuration.

Maximization, on the other hand, exhibits different behavior for
different color-factor configurations.
For configurations of the $\mathcal{P}_1$ and $\mathcal{H}_1$ types,
the leading coefficient is unbounded as $|k|\to\infty$;
consequently, the leading coefficient itself has no finite maximizer
independent of $\eta$.
For a configuration of the $\mathcal{G}_1$ type, the leading
coefficient has a finite global maximum.
Indeed, introducing
\begin{align}
 y\equiv\left(\frac{k-1}{k+1}\right)^2,
\end{align}
the function $\mathcal{G}_1$ can be written as
\begin{align}
 \mathcal{G}_1(k,\eta)
 =
 \frac{2^{16}}{3}
 \cos^{10}\eta\,\sin^4\eta\,
 \frac{
 y^2\left(\sin^2\eta+4y\right)
 }{
 \left(1+4\cos^2\eta\,y\right)^8
 }.
\end{align}
The unique positive stationary point is determined by
\begin{align}
 40\cos^2\eta\,y^2
 +\left(12\sin^2\eta\cos^2\eta-6\right)y
 -\sin^2\eta=0.
\end{align}
Hence, the maximizing values of $k$ depend explicitly on the
initial-correlation parameter $\eta$.
Thus, maximizing the leading coefficient does not select a universal
finite value of $k$ independent of the color-factor configuration
and $\eta$.

\subsection{Series structure of the GGM}
To test the fixed-$k$ asymptotic comparison using an independent
quantitative measure of GTE, we perform the same analysis for the GGM.
In correspondence with the three auxiliary functions introduced for
the concurrence fill, we define
\begin{align}
 \mathcal{P}_2(k,\eta)
 &\equiv
 \frac{(k-1)^2\cos^2\eta}{64},
 \label{eq:P2def}\\
 \mathcal{H}_2(k,\eta)
 &\equiv
 \frac{1}{1024}
 \left[
 4\sin^2\eta
 +9(k-1)^2\cos^2\eta
 \right].
 \label{eq:H2def}
\end{align}
Functional form of $\mathcal{G}_2$ is given in Appendix \ref{app:K0-explicit}. For $0<\eta<\pi/2$, the function $\mathcal{G}_2$ satisfies
\begin{align}
 \mathcal{G}_2(\pm1,\eta)&=0,
 \nonumber\\
 \mathcal{G}_2(k,\eta)&>0,
 \qquad k\neq\pm1.
 \label{eq:G2properties}
\end{align}

\begin{table}[!t]
\begin{minipage}{0.98\textwidth}
 \centering
 \captionof{table}{
 Leading terms in the forward expansion of the GGM for each
 color-factor configuration in gluon scattering and for graviton scattering.
 For $(0,1,-1)$, we consider generic $k\neq\pm1$;
 for the other $\mathcal{P}_2$-type cases, we consider a fixed $k\neq1$. Functional form of $\mathcal{G}_2$ is given in Appendix \ref{app:K0-explicit}.
 }
 \label{tab:GGM-leading-orders}
 \normalsize
 \renewcommand{\arraystretch}{1.2}
 \setlength{\tabcolsep}{7pt}
 \begin{adjustbox}{max width=\linewidth}
 \begin{tabular}{cc}
  \toprule
  $(F_1,F_2,F_3)$ or graviton
  & leading term \\
  \midrule
  $(0,1,-1)$
  & $\mathcal{G}_2(k,\eta)\theta^0$
  \\[2mm]
  $(1,0,1),\ (1,1,0)$
  & $\mathcal{P}_2(k,\eta)\theta^4$
  \\[2mm]
  $(1,2,-1),\ (1,-1,2)$
  & $9\mathcal{P}_2(k,\eta)\theta^4$
  \\[2mm]
  $(2,1,1)$
  & $\mathcal{H}_2(k,\eta)\theta^8$
  \\[2mm]
  graviton
  & $64\mathcal{P}_2(k,\eta)\theta^8$
  \\
  \bottomrule
 \end{tabular}
 \end{adjustbox}
\end{minipage}
\end{table}

Equations~\eqref{eq:Pdef}--\eqref{eq:H1def},
Eqs.~\eqref{eq:P2def}--\eqref{eq:H2def},
and Tables~\ref{tab:leadingorders} and
\ref{tab:GGM-leading-orders} make the correspondence between
the three classes explicit.
The minimization mechanisms of $\mathcal{P}_2$, $\mathcal{G}_2$,
and $\mathcal{H}_2$ are the same as those of
$\mathcal{P}_1$, $\mathcal{G}_1$, and $\mathcal{H}_1$, respectively.

Indeed, both $\mathcal{P}_1$ and $\mathcal{P}_2$ contain explicit
powers of $(k-1)$ and vanish at $k=1$.
Likewise, $\mathcal{H}_1$ and $\mathcal{H}_2$ consist of their values
at $k=1$ together with non-negative contributions depending on
$(k-1)^2$.
For these classes, the leading coefficient at any fixed $k_{*}\neq1$
is therefore larger than that at $k=1$.
Although the powers of $\theta$ differ between the concurrence fill
and the GGM, this difference does not affect the fixed-$k$ ordering.

Similarly, both $\mathcal{G}_1$ and $\mathcal{G}_2$ are positive for
generic fixed $k\neq\pm1$ and vanish at $k=\pm1$.
The leading term of the GGM therefore does not distinguish
$k=1$ from $k=-1$, and the two branches must be compared using
higher-order terms.
The proof of Eq.~\eqref{eq:G2properties}, together with the
demonstration that the apparent $k=-1$ solution is removed once
higher-order terms are included, is given in
Appendix~\ref{app:ggm-analysis}.

The GGM therefore yields the same fixed-$k$ asymptotic selection rule
as the concurrence fill.
Once the comparison point $k_{*}\neq1$ is fixed, the GGM at $k=1$ is
smaller than its value at $k=k_{*}$ at sufficiently small scattering angles.
The GGM thus provides an independent quantitative check of the
concurrence-fill selection rule.
Maximization of the GGM is not considered in this work.

\section{Conclusions}
\label{sec:discussion}
We studied $2+1$-particle gluon and graviton scattering under a one-parameter deformation of the four-point vertex and analyzed the small-angle expansions of the concurrence fill and the GGM of the final state.
To clarify the significance of tripartite correlations, we compared these results with the concurrence of the final state in ordinary two-particle scattering.
We found that, while the bipartite entanglement of the two-particle final state provides no state-independent criterion, the genuine tripartite entanglement of the $2+1$-particle final state singles out the gauge-invariant theory in the forward and backward limits.
In the two-particle setting the same selection was obtained by supplementing the entanglement criterion with the minimization of magic~\cite{Nunez2026Gauge}; here it follows from entanglement alone.

For color-factor configurations with nontrivial $k$ dependence, the local behavior of the concurrence near $k=1$ depends on the initial helicities.
It is minimized for $\ket{++}$ and locally maximized for $\ket{+-}$, whereas it is $k$-independent for certain color-factor configurations.
Bipartite entanglement alone therefore does not determine whether minimization or maximization should serve as the selection criterion.
As a concrete effect of gauge invariance on forward scattering, we also demonstrated that the tree-level equal-helicity pair-flip amplitude $\ket{++}\leftrightarrow\ket{--}$ vanishes at $k=1$ in a particular color-factor class.
Since helicity flips are in general induced at higher orders in perturbation theory, this suppression should be understood as a tree-level effect.

For the concurrence fill, by contrast, the fixed-$k$ asymptotic comparison favors $k=1$ for all initial-correlation parameters $0<\eta<\pi/2$ in Eq.~\eqref{eq:initial}.
This ordering holds at sufficiently small scattering angles for all six gluon color-factor configurations listed in Table~\ref{tab:leadingorders}, as well as for graviton scattering.
For the exceptional value $k=-1$ in the $(0,1,-1)$ configuration, the higher-order terms also favor $k=1$.
The same selection rule holds for the GGM.

These statements do not imply that $k=1$ is the exact global minimizer of the concurrence fill or the GGM at every finite scattering angle.
They establish an asymptotic selection rule in which the comparison point is fixed before the forward or backward limit is taken.

Under the particular four-point vertex deformation adopted here, the gauge-invariant theory in gluon-gluon scattering and the diffeomorphism-invariant theory in graviton-graviton scattering therefore suppress the concurrence fill and the GGM more strongly than any fixed deformation in the forward or backward limit.
For the concurrence, by contrast, the type of extremum depends on the initial helicities, so neither maximization nor minimization provides a state-independent selection principle on its own.
Furthermore, maximizing the leading coefficient of the concurrence fill does not yield a finite value of $k$ that is independent of the initial-correlation parameter $\eta$.

Numerical analyses using two other measures of GTE, the three-$\pi$ entanglement~\cite{OuFan2007} and the genuine multipartite concurrence (GMC)~\cite{Ma2011}, likewise favor $k=1$.
We have not established this selection rule analytically for these two measures and therefore do not include them among the main conclusions of this paper.

\acknowledgments

The author thanks Teppei Kitahara and Kei-Ichi Kondo for helpful discussions and valuable comments.
The author is also grateful to Kazuki Sakurai and Masanori Tanaka for useful discussions.

\appendix
\section{Deformed Yang--Mills helicity amplitudes}
\label{app:amplitudes}

Here, we summarize the independent helicity amplitudes used in this work.
We treat the initial-state particles as incoming external lines and the final-state particles as outgoing external lines, and define the helicity of each particle along its direction of motion.
The same convention is used in Appendix~\ref{app:graviton}.
We use the amplitude conventions of Ref.~\cite{Nunez2026Universality}; see also Refs.~\cite{ManganoParke1991,Dixon1996,ElvangHuang2015} for reviews of helicity amplitudes in gauge theory.
In the gauge-invariant theory ($k=1$), the tree-level amplitudes are
\begin{align}
 \mathcal{M}^{(1)}_{\substack{++\to++\\--\to--}}
 &=2g^2\left[F_3\frac{u-t}{s}
 +F_1\left(2+\frac{u-t}{s}\right)\frac{u}{t}+F_2\left(2-\frac{u-t}{s}\right)\frac{t}{u}\right],\nonumber\\
 \mathcal{M}^{(1)}_{\substack{+-\to+-\\-+\to-+}}
 &=-2g^2\left[F_1\frac{u^2}{ts}+F_2\frac{u}{s}\right],\nonumber\\
 \mathcal{M}^{(1)}_{\substack{+-\to-+\\-+\to+-}}
 &=-2g^2\left[F_1\frac{t}{s}+F_2\frac{t^2}{su}\right].
 \label{eq:M1appendix}
\end{align}
Here, $F_1$, $F_2$, and $F_3$ are the color factors defined in Eq.~\eqref{eq:F123}.
In addition, $\mathcal{M}^{(1)}_{\substack{++\to--\\--\to++}}=\mathcal{M}^{(1)}_{\rm cross}=0$.
Using the definition in Eq.~\eqref{eq:Mcross-definition-main},
the amplitudes after the four-point-vertex deformation can be written as
\begin{align}
 \mathcal{M}_{\substack{++\to++\\--\to--}}&=\mathcal{M}^{(1)}_{\substack{++\to++\\--\to--}}
 +g^2(k-1)\left(F_3\frac{u-t}{s}-F_1\frac{u(2t+u)}{s^2}
 -F_2\frac{t(2u+t)}{s^2}\right),\nonumber\\
 \mathcal{M}_{\substack{+-\to+-\\-+\to-+}}&=\mathcal{M}^{(1)}_{\substack{+-\to+-\\-+\to-+}}
 +g^2(k-1)(F_1+F_2)\left(\frac{u}{s}\right)^2,\nonumber\\
 \mathcal{M}_{\substack{+-\to-+\\-+\to+-}}&=\mathcal{M}^{(1)}_{\substack{+-\to-+\\-+\to+-}}
 +g^2(k-1)(F_1+F_2)\left(\frac{t}{s}\right)^2,\nonumber\\
 \mathcal{M}_{\substack{++\to--\\--\to++}}&=-g^2(k-1)\left(F_3\frac{u-t}{s}+F_1\frac{t(2u+t)}{s^2}
 +F_2\frac{u(2t+u)}{s^2}\right),\nonumber\\
 \mathcal{M}_{\rm cross}&=g^2(k-1)(F_1+F_2)\frac{tu}{s^2}.
 \label{eq:Mdeformedappendix}
\end{align}

\begin{revision}
Up to a common nonzero rescaling and an overall sign, the color-factor configurations realized in massless $SU(3)$ gluon scattering can be classified as follows~\cite{Nunez2026Gauge}.
For amplitudes involving an opposite-helicity initial state, $F_3$ does not contribute, and there are six relations:
\[
 F_1=0,\quad F_2=0,\quad F_1=F_2,\quad F_1=-F_2,\quad
 F_1=2F_2,\quad F_2=2F_1.
\]
For an equal-helicity initial state, there are six relations satisfying the Jacobi identity $F_1-F_2-F_3=0$:
\[
\begin{split}
&F_1=0\neq F_3=-F_2,\qquad
F_2=0\neq F_1=F_3,\qquad
F_3=0\neq F_1=F_2,\\
&F_1=2F_2=2F_3,\qquad
F_2=2F_1=-2F_3,\qquad
F_3=2F_1=-2F_2.
\end{split}
\]
These correspond respectively to the six representatives in Table~\ref{tab:leadingorders}: $(0,1,-1)$, $(1,0,1)$, $(1,1,0)$, $(2,1,1)$, $(1,2,-1)$, and $(1,-1,2)$.
\end{revision}

\section{Graviton helicity amplitudes}
\label{app:graviton}
We follow Refs.~\cite{Nunez2026Gauge,Sannan1986} for the kinematics and for the form of the helicity amplitudes, with the metric convention of Sec.~\ref{sec:amplitudes}.
General relativity is described by the Einstein--Hilbert action
\begin{align}
 S_{\mathrm{GR}}
 =\frac{2}{\kappa^2}
 \int d^4x\,\sqrt{-g}\,R,
 \label{eq:gr-einstein-hilbert-action}
\end{align}
where $g$ is the determinant of the spacetime metric $g_{\mu\nu}$, $R$ is the Ricci scalar, $\kappa^2=32\pi G_{\mathrm{N}}$, and $G_{\mathrm{N}}$ is Newton's constant.

In perturbative gravity, the metric is decomposed as
\begin{align}
 g_{\mu\nu}=\eta_{\mu\nu}+\kappa h_{\mu\nu},
 \label{eq:gr-metric-decomposition}
\end{align}
where $\eta_{\mu\nu}$ is the Minkowski metric introduced in Sec.~\ref{sec:amplitudes} and $h_{\mu\nu}$ is a small perturbation about it that describes a massless spin-$2$ graviton.
The amplitudes below are evaluated in the transverse--traceless gauge, with on-shell kinematics satisfying
\begin{align}
 s+t+u=0.
 \label{eq:gr-onshell-condition}
\end{align}

Summing the contributions from the $s$-, $t$-, and $u$-channel diagrams and the four-point contact vertex, the nonzero full amplitudes in general relativity are
\begin{align}
 \mathcal{M}^{(1)}_{\substack{++\to++\\--\to--}}
 &=-\frac{i\kappa^2}{4}\frac{s^3}{tu},
 \nonumber\\
 \mathcal{M}^{(1)}_{\substack{+-\to+-\\-+\to-+}}
 &=-\frac{i\kappa^2}{4}\frac{u^3}{st},
 \nonumber\\
 \mathcal{M}^{(1)}_{\substack{+-\to-+\\-+\to+-}}
 &=-\frac{i\kappa^2}{4}\frac{t^3}{su}.
 \label{eq:gr-total-amplitudes}
\end{align}
In addition, $\mathcal{M}^{(1)}_{++\to--}=\mathcal{M}^{(1)}_{--\to++}=\mathcal{M}^{(1)}_{\rm cross}=0$.

After the four-point-vertex deformation, the amplitudes are
\begin{align}
 \mathcal{M}_{\substack{++\to++\\--\to--}}
 &=
 -\frac{i\kappa^2}{4}
 \left[
  \frac{s^3}{tu}
  +8(k-1)\frac{t^2u^2}{s^3}
 \right],
 \nonumber\\
 \mathcal{M}_{\substack{+-\to+-\\-+\to-+}}
 &=
 -\frac{i\kappa^2}{4}\frac{u^3}{st},
 \nonumber\\
 \mathcal{M}_{\substack{++\to--\\--\to++}}
 &=
 2i\kappa^2(k-1)\frac{tu}{s^3}
 \left(2t^2+tu+2u^2\right),
 \nonumber\\
 \mathcal{M}_{\substack{+-\to-+\\-+\to+-}}
 &=
 -\frac{i\kappa^2}{4}\frac{t^3}{su},
 \nonumber\\
 \mathcal{M}_{\rm cross}
 &=
 -2i\kappa^2(k-1)\frac{t^2u^2}{s^3}.
 \label{eq:gr-k-deformed-amplitudes}
\end{align}
At $k=1$, the general-relativistic amplitudes in Eq.~\eqref{eq:gr-total-amplitudes} are recovered.

\section{Relation between the forward and backward expansions}
\label{app:forwardandback}

Here, we describe the correspondence between forward and backward scattering.
First, the kinematics under the forward--backward transformation satisfy
\begin{align}
    t(\theta) = -s\sin^2\frac{\theta}{2},\quad u(\theta) = -s\cos^2\frac{\theta}{2},
\end{align}
and hence
\begin{align}
    t(\pi-\theta) = u(\theta),\quad u(\pi-\theta) = t(\theta).
\end{align}
It follows that
\begin{align}
    \theta\to\pi-\theta\quad\Leftrightarrow\quad t\leftrightarrow u.
\end{align}

The color factors also change under the forward--backward transformation.
The operation that maps forward to backward scattering corresponds to interchanging the two final-state gluons,
\begin{align}
    a' \leftrightarrow b'.
\end{align}
Under this interchange,
\begin{align}
    F_1 &\rightarrow f^{ab'c}f^{ba'c} = F_2, \nonumber\\
    F_2 &\rightarrow f^{aa'c}f^{bb'c} = F_1, \nonumber\\
    F_3 &\rightarrow f^{abc}f^{b'a'c} = -f^{abc}f^{a'b'c} = -F_3.
\end{align}
The $\mathcal{M}$ matrix is therefore invariant under the transformation
\begin{align}
    t\leftrightarrow u,\quad (F_1, F_2, F_3) \to (F_2, F_1, -F_3).
\end{align}
Consequently,
\begin{align}
    \text{forward expansion} \to \text{backward expansion}\quad \Leftrightarrow \quad (F_1, F_2, F_3) \to (F_2, F_1, -F_3).
\end{align}
\begin{revision}
Because the concurrence, concurrence fill, and GGM are invariant under interchange of the final-state particles, the fixed-$k$ forward-expansion results carry over to the corresponding backward expansion.
\end{revision}

\section{\textcolor{black}{Vanishing of the equal-helicity pair-flip amplitude in the forward limit}}
\label{app:helicityflip}

\begin{revision}
For a particular color-factor class of gluon $2\to2$ scattering, we show that gauge invariance forces the tree-level equal-helicity pair-flip amplitude to vanish in the forward limit.
\end{revision}
Consider the most general pure initial state of two particles,
\begin{align}
\ket{\Phi_{\rm in}} = c_1\ket{++} +c_2\ket{+-} +c_3\ket{-+} +c_4\ket{--}, \qquad |c_1|^2+|c_2|^2+|c_3|^2+|c_4|^2=1.
\label{eq:general-two-body-initial}
\end{align}
We focus on the nontrivial color-factor class in which the $t$-channel pole vanishes,
\begin{align}
F_1=0, \qquad F_2=-F_3\neq 0.
\label{eq:forward-color-branch}
\end{align}

In this class, the $1/t$ pole vanishes while the $s$-channel contribution remains.
Consequently, a finite $\mathcal{O}(\theta^0)$ amplitude is the leading contribution in the forward limit $\theta\to0$.
This follows directly from Eq.~\eqref{eq:M1appendix}.
Taking the helicity basis in the order
\begin{align}
\left\{ \ket{++},\ket{+-},\ket{-+},\ket{--} \right\},
\end{align}
Writing $\mathcal{M}(k)$ for the matrix with entries
$\mathcal{M}_{\lambda_A\lambda_B\to\lambda_A'\lambda_B'}$ of
Eq.~\eqref{eq:helicity-amplitude-convention} in this basis,
the forward expansion of Eqs.~\eqref{eq:M1appendix} and \eqref{eq:Mdeformedappendix} gives
\begin{align}
\lim_{\theta\to0}\mathcal{M}(k)\equiv \mathcal{M}_0(k)
=g^2
\begin{pmatrix} x & 0 & 0 & z\\ 0 & y & 0 & 0\\ 0 & 0 & y & 0\\ z & 0 & 0 & x
\end{pmatrix},
\label{eq:forward-helicity-matrix-general}
\end{align}
where
\begin{align}
x=-(k+1)F_3, \qquad y=(k+1)F_2, \qquad z=(k-1)(F_3-F_2).
\label{eq:xyz-two-body}
\end{align}
Defining $F\equiv F_2=-F_3\neq0$, we obtain $x=y=(k+1)F$ and $z=-2(k-1)F$.
The forward-scattering matrix is therefore
\begin{align}
\mathcal{M}_0(k) = g^2F \begin{pmatrix} k+1 & 0 & 0 & -2(k-1)\\ 0 & k+1 & 0 & 0\\ 0 & 0 & k+1 & 0\\ -2(k-1) & 0 & 0 & k+1
\end{pmatrix}.
\label{eq:forward-helicity-matrix-branch}
\end{align}
Because the diagonal entries represent processes that do not change helicity during scattering, the only helicity-changing amplitude at $\mathcal{O}(\theta^0)$ is
\begin{align}
\mathcal{M}_{++\to--} = \mathcal{M}_{--\to++} = -2g^2(k-1)F.
\label{eq:pair-helicity-flip}
\end{align}
Thus, the helicity flip relevant to this color-factor class is
\begin{align}
\ket{++}\longleftrightarrow\ket{--}.
\end{align}
The nonzero pair-flip amplitude at $k\neq1$ accounts for the forward-scattering entanglement found in Sec.~\ref{subsec:seriesconcurrenceRR} for the initial state $\ket{++}$ and the color-factor configuration $(F_1,F_2,F_3)=(0,1,-1)$.
In this class, the pair-flip amplitude vanishes at tree level only in the gauge-invariant theory ($k=1$).

\section{Structure of $\mathcal{G}_1$}
\label{app:G1structure}

\begin{revision}
For the color-factor configuration $(0,1,-1)$, this appendix compares $\CF$ at fixed values of $k$ in the small-angle limit.
We define the forward expansion of its fourth power by
\begin{align}
 \CF^4(k,\theta,\eta;(F_1,F_2,F_3))
 =\sum_{n=0}^{\infty}
 c_n(k,\eta;(F_1,F_2,F_3))\,\theta^n.
 \label{eq:CF4-coefficient-definition}
\end{align}
From Table~\ref{tab:leadingorders} and Eq.~\eqref{eq:G1def},
\begin{align}
 \mathcal{G}_1(k,\eta)=c_0(k,\eta;(0,1,-1)).
\end{align}
For $0<\eta<\pi/2$, $\mathcal{G}_1(k,\eta)\ge0$, and its only zeros are $k=\pm1$.
Thus, comparison with any fixed $k_{*}\neq\pm1$ is already determined at leading order, leaving only the two fixed exceptional branches $k=1$ and $k=-1$.
We compare these two branches at the first order at which their values differ.

Terms through order $\theta^{14}$ do not distinguish the two branches.
The first difference appears at order $\theta^{16}$, and the difference between the corresponding coefficients is
\begin{align}
&c_{16}\bigl(k=1,\eta;(0,1,-1)\bigr)
-c_{16}\bigl(k=-1,\eta;(0,1,-1)\bigr)
\nonumber\\
&\quad =
-\frac{\sec^4\eta\,\tan^4\eta}{25165824}
\Bigl[
374834
+478832\cos(2\eta)
+106178\cos(4\eta)
\nonumber\\
&\hspace{42mm}
+352\cos(6\eta)
+8\cos(8\eta)
-160\cos(10\eta)
-83\cos(12\eta)
\nonumber\\
&\hspace{42mm}
+8\cos(14\eta)
+22\cos(16\eta)
+8\cos(18\eta)
+\cos(20\eta)
\Bigr].
\label{eq:G1-c16-difference}
\end{align}
We now determine the sign of the right-hand side of Eq.~\eqref{eq:G1-c16-difference}.
Denote the expression in square brackets by $\mathcal{B}(\eta)$ and define
\begin{align}
 x\equiv\cos^2\eta.
 \label{eq:G1-y-definition}
\end{align}
For $0<\eta<\pi/2$, one has $0<x<1$.
Expressing $\mathcal{B}(\eta)$ as a polynomial in $x$, we obtain
\begin{align}
 \mathcal{B}(\eta)
 =
 1920
 +111360x
 +846720x^2
 -524288x^7(1-x)^3.
 \label{eq:G1-bracket-polynomial}
\end{align}
Since $0<x<1$ implies
\begin{align}
 0<x^7(1-x)^3\leq x^2,
\end{align}
it follows that
\begin{align}
 \mathcal{B}(\eta)
 &\geq
 1920
 +111360x
 +(846720-524288)x^2
 \nonumber\\
 &=
 1920
 +111360x
 +322432x^2
 >0.
 \label{eq:G1-bracket-positive}
\end{align}
Applying this result to Eq.~\eqref{eq:G1-c16-difference}, we find
\begin{align}
 c_{16}\bigl(k=1,\eta;(0,1,-1)\bigr)
 -
 c_{16}\bigl(k=-1,\eta;(0,1,-1)\bigr)
 <0.
\end{align}

Therefore,
\begin{align}
 \CF^4&(k=1,\theta,\eta;(0,1,-1))
 -\CF^4(k=-1,\theta,\eta;(0,1,-1))
 \nonumber\\
 &=
 \left[
 c_{16}\bigl(1,\eta;(0,1,-1)\bigr)
 -c_{16}\bigl(-1,\eta;(0,1,-1)\bigr)
 \right]\theta^{16}+\mathcal{O}(\theta^{18})<0.
 \label{eq:G1-branch-comparison}
\end{align}
for sufficiently small $\theta$.
The same ordering also holds for $\CF$.
Thus, among the two fixed branches, $k=1$ gives the smaller value at sufficiently small but nonzero forward angles.
\end{revision}

\section{Analytic structure of the GGM for $(F_1,F_2,F_3)=(0,1,-1)$}
\label{app:ggm-analysis}
\subsection{Explicit form of $\mathcal{G}_2(k,\eta)$}
\label{app:K0-explicit}
In this subsection, we present the explicit form of $\mathcal{G}_2(k,\eta)$ introduced in Table~\ref{tab:GGM-leading-orders}.
To simplify the expressions, we define
\begin{align}
 p_k\equiv(k+1)^2,\qquad
 d_k\equiv4(k-1)^2,\qquad
 \mathcal{N}_0(k,\eta)
 \equiv
 3-2k+3k^2
 +2(k-1)^2\cos2\eta.
 \label{eq:K0-forward-normalization}
\end{align}
Here, $\mathcal{N}_0(k,\eta)$ is the squared norm of the unnormalized final state.

Writing explicitly the GGM candidates associated with the three bipartitions $A|BC$, $B|AC$, and $C|AB$ as functions of $k$ and $\eta$, we obtain
\begin{align}
 \mathcal{G}_A(k,\eta)
 &=
 \frac{1}{2}
 \left[
 1-
 \frac{
 \left|
 1-6k+k^2
 +2(k-1)^2\cos2\eta
 \right|
 }{
 \mathcal{N}_0(k,\eta)
 }
 \right],
 \label{eq:K0-expanded-A}\\
 \mathcal{G}_B(k,\eta)
 &=
 \frac{1}{2}
 \left[
 1-
 \frac{
 \left|
 2(k-1)^2
 +(1-6k+k^2)\cos2\eta
 \right|
 }{
 \mathcal{N}_0(k,\eta)
 }
 \right],
 \label{eq:K0-expanded-B}\\
 \mathcal{G}_C(k,\eta)
 &=
 \frac{1}{2}
 \left[
 1-
 \frac{
 \left|
 2(k-1)^2
 +(3-2k+3k^2)\cos2\eta
 \right|
 }{
 \mathcal{N}_0(k,\eta)
 }
 \right].
 \label{eq:K0-expanded-C}
\end{align}
Here $\mathcal{G}_X(k,\eta)\equiv\lim_{\theta\to0}g_X$ $(X=A,B,C)$,
with $g_X$ defined in Eq.~\eqref{eq:GGM-branches}.
Thus, $\mathcal{G}_2(k,\eta)$ is given by
\begin{align}
 \mathcal{G}_2(k,\eta)
 =
 \min\left\{
 \mathcal{G}_A(k,\eta),
 \mathcal{G}_B(k,\eta),
 \mathcal{G}_C(k,\eta)
 \right\}.
 \label{eq:K0-minimum-form}
\end{align}

We next specify the regions in which each candidate is selected;
$\Omega_X$ denotes the condition for $\mathcal{G}_2=\mathcal{G}_X$ $(X=A,B,C)$.
The condition for selecting the $A|BC$ branch is
\begin{align}
 \Omega_A
 \Longleftrightarrow{}&
 \Biggl[
 0<\eta<\frac{\pi}{4}
 \ \land\
 \Bigl\{
 d_k\cos^2\eta<p_k\sin^2\eta
 {}\lor
 \bigl(
 d_k\cos^2\eta=p_k\sin^2\eta
 \land k<1
 \bigr)
 \Bigr\}
 \Biggr]
 \nonumber\\
 &{}\lor
 \left[
 \eta=\frac{\pi}{4}
 \ \land\
 \frac{1}{3}\leq k<3
 \right]
 \nonumber\\
 &{}\lor
 \Biggl[
 \frac{\pi}{4}<\eta<\frac{\pi}{2}
 \ \land\
 \Bigl\{
 d_k<p_k
 {}\lor
 \bigl(d_k=p_k\land k<1\bigr)
 \Bigr\}
 \Biggr].
 \label{eq:K0-condition-A}
\end{align}

\begin{figure}[!t]
  \centering
  \IfFileExists{
    G2_vs_k.pdf
  }{%
    \includegraphics[width=0.6\linewidth,keepaspectratio]{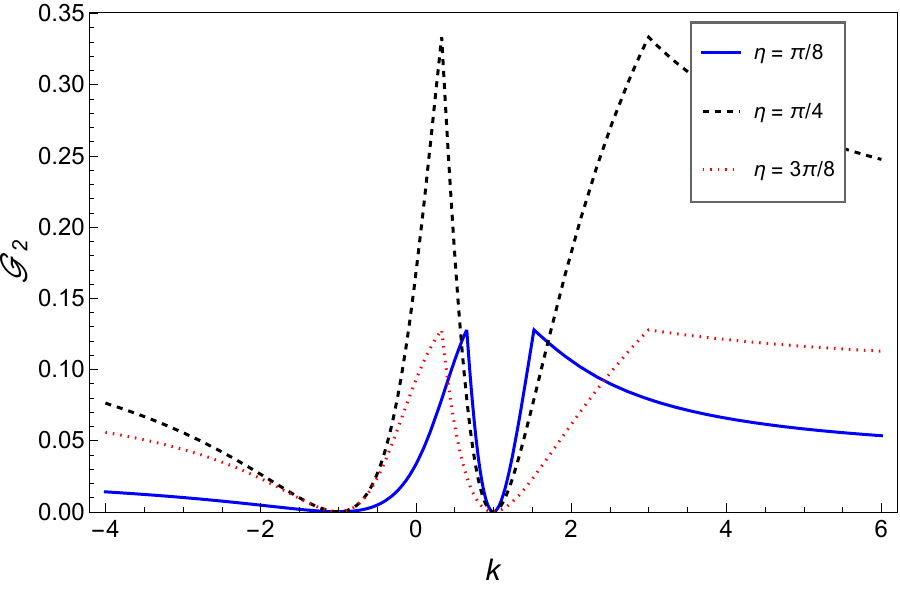}
  }{%
    \fbox{%
      \parbox[c][32mm][c]{0.6\linewidth}{%
        \centering
        Place the figure file in the same directory
      }%
    }%
  }
  \caption{The function $\mathcal{G}_2(k,\eta)$ of Eq.~\eqref{eq:K0-minimum-form} as a function of $k$ for several representative values of $\eta$.
  It is non-negative and vanishes only at $k=\pm1$.}
  \label{fig:G2plot}
\end{figure}

Similarly, the condition for selecting the $B|AC$ branch is
\begin{align}
 \Omega_B
 \Longleftrightarrow{}&
 \Biggl[
 \frac{\pi}{4}<\eta<\frac{\pi}{2}
 \ \land\
 \Bigl\{
 d_k>p_k
 {}\lor
 \bigl(d_k=p_k\land k>1\bigr)
 \Bigr\}
 \Biggr]
 \nonumber\\
 &{}\lor
 \left[
 \eta=\frac{\pi}{4}
 \ \land\
 \bigl(k<-1\lor k>3\bigr)
 \right],
 \label{eq:K0-condition-B}
\end{align}
while the condition for selecting the $C|AB$ branch is
\begin{align}
 \Omega_C
 \Longleftrightarrow{}&
 \Biggl[
 0<\eta<\frac{\pi}{4}
 \ \land\
 \Bigl\{
 d_k\cos^2\eta>p_k\sin^2\eta
 {}\lor
 \bigl(
 d_k\cos^2\eta=p_k\sin^2\eta
 \land k>1
 \bigr)
 \Bigr\}
 \Biggr]
 \nonumber\\
 &{}\lor
 \left[
 \eta=\frac{\pi}{4}
 \ \land\
 \left(
 -1\leq k<\frac{1}{3}
 \ \lor\
 k=3
 \right)
 \right].
 \label{eq:K0-condition-C}
\end{align}
Because multiple candidates coincide at branch boundaries, assigning a boundary point to any adjacent region does not affect the value of $\mathcal{G}_2(k,\eta)$ itself.
It therefore follows from Eq.~\eqref{eq:K0-minimum-form} that
\begin{align}
 \mathcal{G}_2(\pm1,\eta)&=0,
 \nonumber\\
 \mathcal{G}_2(k,\eta)&>0,
 \qquad
 k\neq\pm1.
 \label{eq:K0-positivity}
\end{align}
This establishes that the $\theta^0$ contribution to the GGM for generic fixed $k$ vanishes only at $k=\pm1$.
Figure~\ref{fig:G2plot} illustrates this behavior.

\subsection{Higher-order series structure}

We now show that, at sufficiently small scattering angles, the GGM at $k=1$ is smaller than its value at any fixed $k\neq1$.
Equation~\eqref{eq:K0-positivity} determines the comparison with every fixed $k\neq\pm1$ at leading order, so we need only compare $k=1$ and $k=-1$.
First, substituting the values of the $\mathcal{M}$ matrix into Eq.~\eqref{eq:outstate}, the unnormalized final state in the limit $\theta\to0$ is
\begin{align}
 \ket{\widetilde{\Psi}_{\mathrm{out},0}}
 =
 (k+1)\cos\eta\,\ket{+++}
 -2(k-1)\cos\eta\,\ket{--+}+
 (k+1)\sin\eta\,\ket{+--}.
 \label{eq:GGM-01m1-forward-state}
\end{align}
To express the higher-order comparison compactly, we use the variable $q$
of Eq.~\eqref{eq:q-definition} and write $g_X(k,q,\eta)$ for the branches of
Eq.~\eqref{eq:GGM-branches} evaluated for the configuration $(0,1,-1)$.
At $k=1$, the three branches are
\begin{align}
 g_A(1,q,\eta)
 &=
 \sin^2\eta\,q^4+\mathcal{O}(q^5),
 \nonumber\\
 g_B(1,q,\eta)
 &=
 \frac{1-|\cos2\eta|}{2}
 +\mathcal{O}(q),
 \nonumber\\
 g_C(1,q,\eta)
 &=
 \frac{1-|\cos2\eta|}{2}
 +\mathcal{O}(q).
 \label{eq:GGM-01m1-k1-branches}
\end{align}
Thus, for sufficiently small $q$, the $A|BC$ branch is selected, yielding
\begin{align}
 \mathcal{E}_{\mathrm{GGM},(0,1,-1)}(1,q,\eta)
 =
 \sin^2\eta\,q^4+\mathcal{O}(q^5).
 \label{eq:GGM-01m1-k1}
\end{align}

At $k=-1$, on the other hand, the forward expansions are
\begin{align}
 g_A(-1,q,\eta)
 &=
 \frac{25+2\tan^2\eta}{4}\,q^2
 +\mathcal{O}(q^3),
 \nonumber\\
 g_B(-1,q,\eta)
 &=
 \frac{25+\tan^2\eta}{4}\,q^2
 +\mathcal{O}(q^3),
 \nonumber\\
 g_C(-1,q,\eta)
 &=
 \frac{\tan^2\eta}{2}\,q^2
 +\mathcal{O}(q^3).
 \label{eq:GGM-01m1-kminus1-branches}
\end{align}
Thus, $g_A$ is not the smallest branch, and
\begin{align}
 \mathcal{E}_{\mathrm{GGM},(0,1,-1)}(-1,q,\eta)
 =
 D_-(\eta)q^2+\mathcal{O}(q^3),
 \label{eq:GGM-01m1-kminus1}
\end{align}
where
\begin{align}
 D_-(\eta)
 \equiv
 \begin{cases}
  \dfrac{\tan^2\eta}{2},
  & \tan^2\eta\leq25,
  \\[2mm]
  \dfrac{25+\tan^2\eta}{4},
  & \tan^2\eta>25
 \end{cases}.
 \label{eq:GGM-Dminus}
\end{align}
That is, the $C|AB$ branch is selected for $\tan^2\eta<25$, the $B|AC$ branch is selected for $\tan^2\eta>25$, and the two branches coincide at $\tan^2\eta=25$.

In either case, $D_-(\eta)>0$, and hence
\begin{align}
 \mathcal{E}_{\mathrm{GGM},(0,1,-1)}(1,q,\eta)
 &=\mathcal{O}(q^4),
 \nonumber\\
 \mathcal{E}_{\mathrm{GGM},(0,1,-1)}(-1,q,\eta)
 &=\mathcal{O}(q^2).
 \label{eq:GGM-01m1-orders}
\end{align}
Therefore, for sufficiently small but finite $q>0$,
\begin{align}
 \mathcal{E}_{\mathrm{GGM},(0,1,-1)}(1,q,\eta)
 <
 \mathcal{E}_{\mathrm{GGM},(0,1,-1)}(-1,q,\eta),
 \label{eq:GGM-01m1-selection}
\end{align}
so $k=1$ is favored over $k=-1$.

\bibliographystyle{JHEP}
\bibliography{refs}

\end{document}